\documentclass[prl,twocolumn,superscriptaddress,preprintnumbers,amsmath,amssymb]{revtex4}

\usepackage[dvips]{graphicx}
\usepackage{amsfonts}
\usepackage{lmodern}
\usepackage{textcomp}
\usepackage[T1]{fontenc}
\usepackage{gensymb} 
\usepackage[seperr, load={}]{siunitx}
\usepackage{bm}
\usepackage{color}
\usepackage{ulem} 

\newcommand*\coloron{(Color online)\ }
\newcommand*{\figref}[1]{Fig.~\ref{#1}}
\newcommand*\paper{Letter}

\newcommand{\mysec}[1]{\smallskip\textit{#1}} 
\newcommand*\supplmat{\cite{Suppl_Mat}}

\newcommand*{\ket}[1]{|{#1}\rangle}

\newcommand*{\mean}[1]{\mathinner{\langle{#1}\rangle}}

\newcommand*{\xbraket}[3]{\mathinner{\langle{#1}|{#2}|{#3}\rangle}}

\newcommand*{\Fst}{\Psi_\alpha} 
\newcommand*{\Fmd}{\varphi_\alpha} 
\newcommand*{\Fqe}{\epsilon_\alpha} 
\newcommand*{\Fnoise}{X_{\alpha,\beta,k}} 
\newcommand*{\Fgap}{\mathcal {E}} 
\newcommand*{\Dabk}{\Delta_{\alpha\beta k}} 
\newcommand*{\nth}{n_{\rm th}} 
\newcommand*{\al}{\alpha} 
\newcommand*{\be}{\beta} 
\newcommand*{\ga}{\gamma} 
\newcommand*{\de}{\delta} 
\newcommand*{\Jmax}{J_{\rm max}} 
\newcommand*{\Jmin}{J_{\rm min}} 
\newcommand*{\rhost}{\rho_{\rm st}} 

\begin{document}

\title{Environment-governed dynamics in driven quantum systems}

\author{S. Gasparinetti}
	\email{simone.gasparinetti@aalto.fi}
	\affiliation{Low Temperature Laboratory (OVLL), Aalto University, P.O. Box
	15100, FI-00076 Aalto, Finland}
\author{P. Solinas}
 	\affiliation{Low Temperature Laboratory (OVLL), Aalto University, P.O. Box
	15100, FI-00076 Aalto, Finland}
	\affiliation{COMP Centre of Excellence, Department of Applied Physics, Aalto University School of Science,
P.O. Box 11000, 00076 Aalto, Finland
}
\author{S. Pugnetti}
	\affiliation{NEST, Scuola Normale Superiore and Istituto Nanoscienze-CNR,
I-56126 Pisa, Italy}
\author{R. Fazio}
	\affiliation{NEST, Scuola Normale Superiore and Istituto Nanoscienze-CNR,
I-56126 Pisa, Italy}
\author{J. P. Pekola}
	\affiliation{Low Temperature Laboratory (OVLL), Aalto University, P.O. Box
	15100, FI-00076 Aalto, Finland}
\date{\today}

\begin{abstract}
We show that the dynamics of a driven quantum system weakly coupled to the environment can exhibit two
distinct regimes. While the relaxation basis is usually determined by the system+drive Hamiltonian (system-governed dynamics),
we find that under certain conditions it is determined by specific features of the environment, such as, the form of the coupling operator
(environment-governed dynamics).
We provide an effective coupling parameter describing the transition between the two regimes
and discuss how to observe the transition in a superconducting charge pump.
\end{abstract}

\maketitle

\mysec{Introduction.}
Understanding how quantum systems interact with the environment \cite{Breuer2007}
is of paramount importance in quantum information science.
While unveiling how the classical world
emerges from the quantum one \cite{Zurek2003},
it can also lead to a better protection against decoherence
effects on the way towards the realization of a quantum computer
\cite{Nielsen2000}.

A standard approach to the dynamics of open quantum systems boils the problem
down to the measurement of decoherence rates, distinguishing between coherence
loss, or dephasing, and relaxation.
While this approach has successfully described a variety of quantum systems, it
only offers a limited insight into the dynamics of decoherence.
A promising line of work developed in the last decade exploits the possibility of
coupling the system to an engineered reservoir
\cite{Myatt2000,Kielpinski2001,Weimer2010,Solinas2010,Barreiro2011,Diehl2011,Liu2011,Solinas2012,Murch2012}.

As new and more accurate ways are found of harnessing the dynamic evolution of
quantum systems, it becomes increasingly important to understand how the
interaction with the environment is affected by a time-dependent modulation of
the system parameters. Indeed, the study of dissipation in driven quantum systems
is a long-established topic \cite{Grifoni1998} that keeps finding new applications to
quantum pumping \cite{Pekola2010,Russomanno2011,Pellegrini2011}, quantum computation
\cite{Diehl2008,Ferron2012} and possibly even biological systems \cite{Galve2010,Cai2010a}.

In this \paper, we consider a periodically driven quantum system in the presence
of a weakly coupled environment. We show that under certain conditions
decoherence takes place in a preferred basis determined by specific features of
the environment, such as, the type of noise, rather than of the system. We label
this unusual regime as environment-governed dynamics (EGD), as opposed to the
more familiar system-governed dynamics (SGD).
We introduce an effective coupling parameter that presides over the transition
between SGD and EGD. This parameter can be tuned by changing the properties of
the drive. Our analysis is general and applies to optical and
solid-state systems alike.
As a relevant example, we propose to observe the transition in a superconducting charge pump
\cite{Niskanen2003,Niskanen2005}. In this system the transition is controlled by an accessible
experimental parameter and can be explored by measuring the pumped charge.

\mysec{Floquet-Born-Markov master equation.}
We consider a quantum system whose unitary evolution is governed by a periodic
Hamiltonian $H$, so that $H(t) = H(t+\tau)$, where $\tau$ is the period.
According to Floquet theorem, the Schr\"odinger equation admits solutions
(Floquet states) of the form $\ket{\Fst(t)}=e^{-i\Fqe t/\hbar}\ket{\Fmd(t)}\ ,$
where the Floquet mode $\ket{\Fmd(t)}$ satisfies
$\ket{\Fmd(t+\tau)}=\ket{\Fmd(t)}$ and $\Fqe$ is its corresponding quasienergy.
Quasienergies and their associated modes are defined up to the translation $\Fqe
\to \Fqe + \hbar \Omega$, where $\Omega=2\pi/\tau$. As such, all quasienergies
can be mapped into the first Brillouin zone
$[-\frac12\hbar\Omega,\frac12\hbar\Omega]$.
The Hamiltonian describing the system and its environment is given by
$H_{tot}=H(t)+H_E+H_{SE}$, where $H_E$ describes the environmental degrees of
freedom and $H_{SE}$ is the interaction term, that we assume of the form
$H_{SE}= g~A \otimes E$, where $g$ is an adimensional coupling constant and $A$
and $E$ are operators acting in the Hilbert space of the system and the
environment, respectively.

The general procedure
for deriving the master equation (ME) in the Floquet basis and in the Born-Markov approximation is
outlined in Refs.~\onlinecite{Blumel1991,Grifoni1998}.
The resulting superoperator describing the time evolution of the density matrix is expressed a series
of time-independent coefficients multiplied by phase factors of the form $e^{i\Delta_{\alpha,\beta,k}-i\Delta_{\gamma,\delta,k'}}$,
with $\Delta_{\alpha,\beta,k}=\epsilon_\alpha-\epsilon_\beta+k\Omega$.
Starting from the general expression, we then perform a partial secular approximation (PSA).
This amounts to neglecting all terms with $k\neq k'$ while keeping those with $k=k'$ also when
$\epsilon_\alpha \neq \epsilon_\beta$.
A residual time dependence due to terms oscillating like $e^{i
(\epsilon_\alpha-\epsilon_\beta) t}$ disappears when passing from the basis of
Floquet states to that of Floquet modes.
Our result, written in the basis of Floquet modes and in the Schr\"odinger
picture, reads:
\begin{equation}
\dot{\rho}_{\alpha \beta} =- i (\epsilon_\alpha-\epsilon_\beta) \rho_{\alpha \beta} + \sum_{\gamma ,\delta } \rho _{\gamma ,\delta } \mathcal{R}_{\alpha
   ,\beta ,\gamma ,\delta }
   \label{eq:blum}
\end{equation}
where 
\begin{equation}
\begin{split} \label{eq:Rdef}
\mathcal{R}_{\al,\be,\ga,\de}=
\Gamma^+{}_{\al,\ga,\be,\de}
+\Gamma ^-{}_{\al,\ga,\be,\de} \\
- \delta _{\de,\be} \sum_\mu \Gamma^+{}_{\mu,\ga,\mu,\al}
- \delta_{\ga,\al} \sum_\mu \Gamma ^-{}_{\mu,\be,\mu,\de}
   \end{split}
\end{equation}
 and
\begin{subequations}\label{eq:Gammas}
   \begin{gather}
   \Gamma^+{}_{\al,\be,\ga,\de}=
   \sum_k \left. S(\Delta_{\al,\be,k})
   X_{\al,\be,k} \left(X_{\ga,\de,k}\right)^* \right. \label{eq:Gammapl}\ , \\ 
   \Gamma^-{}_{\alpha,\beta,\gamma,\delta} =
   \sum_k \left. S(\Delta_{\ga,\de,k})
    X_{\al,\be,k} \left(X_{\ga,\de,k}\right)^* \right. \label{eq:Gammamn}\ .
\end{gather}
\end{subequations}
We have introduced the following quantities:
\begin{subequations}\label{eq:ME_defs}
\begin{align*}
S(\omega) &= \theta(\omega) J(\omega) \nth(\omega)
+\theta(-\omega)J(-\omega) \left[1+\nth(-\omega)\right] \ , \\
\Dabk &= \epsilon_\alpha-\epsilon_\beta+k\Omega\ , \\
X_{\alpha\beta k} &= \int_0^\tau dt e^{-ik\Omega t}
\xbraket{\varphi_\alpha}{A}{\varphi_\beta}\ ,
\end{align*}
\end{subequations}
where $\theta(\omega)$ is the Heaviside function, $J(\omega)$ the spectral
density of the bath and $\nth(\omega)$ is the Bose-Einstein distribution.

A few remarks are in order.
First, \eqref{eq:blum} cannot be cast into a Pauli ME \cite{Blum2012}, i.e., the
equations for the diagonal and off-diagonal terms are still coupled.
This is due to the fact that we performed a PSA instead of a full secular
approximation; 
we expect the additional terms to become important close to
degeneracies in the Floquet spectrum \cite{Dittrich1993,Oelschlagel1993}.
The PSA itself is justified provided the drive period is much faster than the
decoherence time. This may not be true in the adiabatic limit, where other terms
should instead be retained \cite{Kamleitner2011}.
We have numerically checked the validity of the PSA in the present case by
including more oscillating terms in the ME and comparing the results; the
agreement is excellent.

Equation \eqref{eq:blum} is formally akin to that of an undriven system, with
quasienergies playing the role of ordinary eigenenergies and effective rates
given by \eqref{eq:Gammas}. The presence of a drive manifests itself in the sum
over $k$ in \eqref{eq:Gammas}, allowing the system to exchange energy with the
environment in any of the amounts $\Delta_{\al,\be,k}$. The magnitude of this
exchange is determined by the noise matrix elements $\Fnoise$, telling whether
the Floquet modes possess those energies and to what extent the coupling operator
$A$ allows the energy transfer to take place. Note that while
each individual term in the sums \eqref{eq:Gammas} satisfies the detailed
balance, the overall rates $\Gamma^\pm$ in general do not.

\mysec{Environment-governed dynamics.} Equation \eqref{eq:blum} suggests that the
dynamics of the coherences is determined by a
competition between two terms. Let us now focus our attention on two states,
$\alpha$ and $\beta$.
For $\alpha\neq\beta$, the first term in \eqref{eq:blum} is the Floquet energy gap
$\Fgap=\epsilon_\alpha-\epsilon_\beta$, stemming from the nondissipative
dynamics of the driven quantum system. The second term describes the effect of
dissipation.
For sufficiently small values of $\Fgap$, that is, close to a degeneracy in the
Floquet spectrum, the dissipative term can dominate in \eqref{eq:blum}. As a
result, the dynamics of states $\alpha$, $\beta$ is
strongly affected by the environment even if the system and the environment are
only weakly coupled, i.e., $g \ll 1$.
This is due to the presence of a nearly resonant driving field introducing an
energy scale, the Floquet gap, that can be much smaller than those of the
undriven system.

Let us now estimate the magnitude of the rates $\Gamma^\pm$ in
\eqref{eq:Gammas}. We consider an Ohmic environment, whose spectral density is
given by $J(\omega)  = a \omega f_c(\omega/\omega_c)$, where $a$ is a
dimensionless constant, $f_c$ a cutoff function and $\omega_c$ the cutoff
frequency. In the limit $\Omega \ll \omega_c$ our results are independent of
$\omega_c$ and the explicit form of $f_c$.
As the sum in \eqref{eq:Gammas} runs over $k$, we may expect the dominant
contributions to come from high-frequency modes (large $k$) such that $X_{\alpha
\beta k}$ does not vanish and $k\Omega < \omega_c$.
These contributions are of order $g^2(\Fgap +  k \Omega)\approx k g^2 \Omega$.

We define an effective coupling parameter as
\begin{equation}\label{eq:chi}
\chi=g^2 \Omega /\Fgap\ .
\end{equation}
If $\chi \ll 1$ we are in the SGD regime. This corresponds to a decoherence time
much longer than
the period of
Rabi oscillations between Floquet modes.
By contrast, when $\chi \gtrsim 1$ we enter the EGD regime. This
regime is realized close to a (quasi)degeneracy of the Floquet spectrum, where
Rabi oscillations between Floquet modes are slower than the decoherence
time.
We remark that the coupling parameter $\chi$ is controlled by the Floquet
gap $\Fgap$, which can be tuned by changing the drive parameters.
To appreciate the key role played by the drive, we observe that an undriven system
with an arbitrarily small energy gap cannot exhibit EGD.
This is due to the fact that for any Ohmic and super-Ohmic environment the transition rates vanish
for a vanishing energy gap.

\begin{figure}
    \begin{center}
   \includegraphics{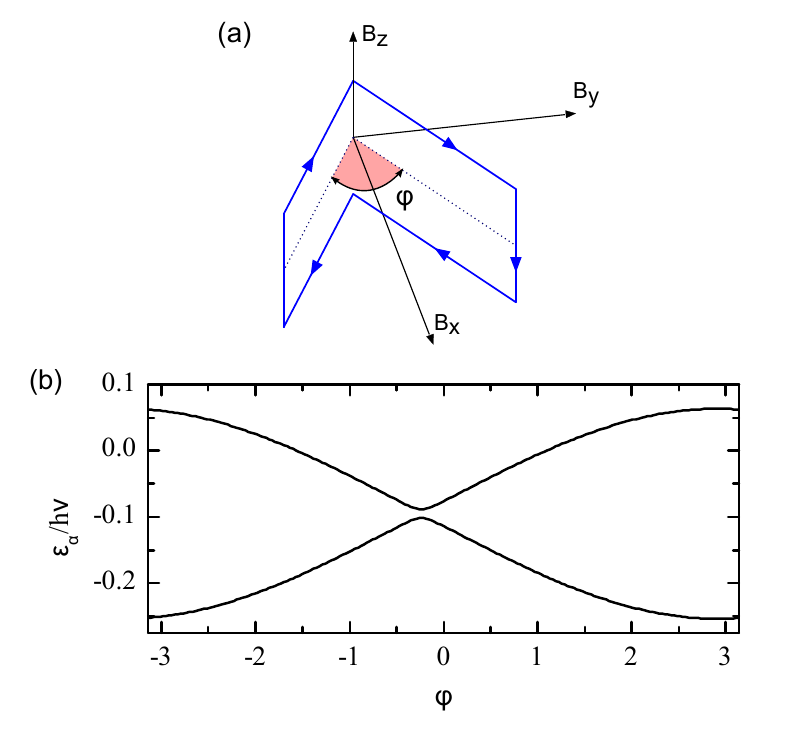}
   \end{center}
    \caption{\coloron A model driven two-level system.
    (a) Due to the driving field, the tip of the effective magnetic field $\vec B$ draws a closed loop 
    in pseudospin space. The explicit time dependence of $\vec B$ can be obtained by that of the parameters $J_L$,
    $J_R$ and $n_g$, provided in \figref{fig:sluice}(b), with the help of the expressions given in the text.
    The solid angle spanned by $\vec B$ is controlled by the angle $\varphi$, that we regard as a tunable parameter. 
    (b) Quasienergy spectrum as a function of $\varphi$. A weakly avoided quasienergy crossing
    occurs at $\varphi_c=-0.26$.}
    \label{fig:Floquet}
\end{figure}

\mysec{SGD to EGD transition.}
As the foregoing discussion highlights, a transition between SGD and EGD is to be expected
every time (i) a degeneracy or quasidegeneracy is encountered in the Floquet spectrum, and (ii)
the noise operator actively couples the Floquet states. These conditions are quite general and may be found in
a variety of systems; even a simple quantum bit driven by a monochromatic drive can exhibit EGD.
Furthermore, environments with a more structured density of states, such as those found in cavity quantum
electrodynamics architectures, may offer additional insight into the EGD regime.

We now provide an explicit example of SGD to EGD transition in a driven
two-level system.
Using a pseudospin formalism, we write the system Hamiltonian as
$H(t)=\vec{\sigma}\cdot \vec{B}(t)$, where $\vec{\sigma}= \{\sigma_x, \sigma_y,
\sigma_z\}$ are the Pauli operators and $\vec{B}(t)$ an effective magnetic
field.
We consider a drive that modulates $\vec{B}$ along the loop shown in
\figref{fig:Floquet}(a).
This drive is a realization of Landau-Zener-St\"uckelberg interference
\cite{Shevchenko2010} with geometric phases \cite{Gasparinetti2011a}.
Its geometric properties provide a convenient way of tailoring the Floquet
spectrum. In particular, we study the effects of varying the angle $\varphi$
[see \figref{fig:Floquet}(a)], which determines the solid angle spanned
by $\vec{B}$ in the pseudospin space during a drive period.
In \figref{fig:Floquet}(b) we plot the quasienergy spectrum of the system versus
$\varphi$.
The plot is obtained by numerically solving the Schr\"odinger equation for the
evolution operator generated by $H$.
The quasienergy gap $\Fgap$ sharply decreases near $\varphi=\varphi_c$, where a
weakly avoided quasienergy crossing occurs.
Close to $\varphi_c$, we thus expect EGD to be attained.

We characterize the transition from SGD to EGD by studying the quasistationary
state approached by the driven system in the presence of dissipation.
For the purpose of illustration, we consider a zero-temperature environment.
\footnote{The finite-temperature behavior is qualitatively similar to that reported as long as
the condition $k_B T\ll \mean{\Delta E}$, where $\mean{\Delta E}$ is the mean energy gap.} 
We write the quasistationary density matrix $\rhost$ as $\rhost  = 1/2(\openone
+ \vec{n} \cdot \vec{\Sigma})$ where $\vec \Sigma$ is the vector of Pauli operators in
the Floquet-mode basis. In this way, the residual coherence between Floquet
states is associated to the quantity $n_\perp=\sqrt{n_x^2+n_y^2}$.

\begin{figure}
    \begin{center}
    \includegraphics[width=\linewidth]{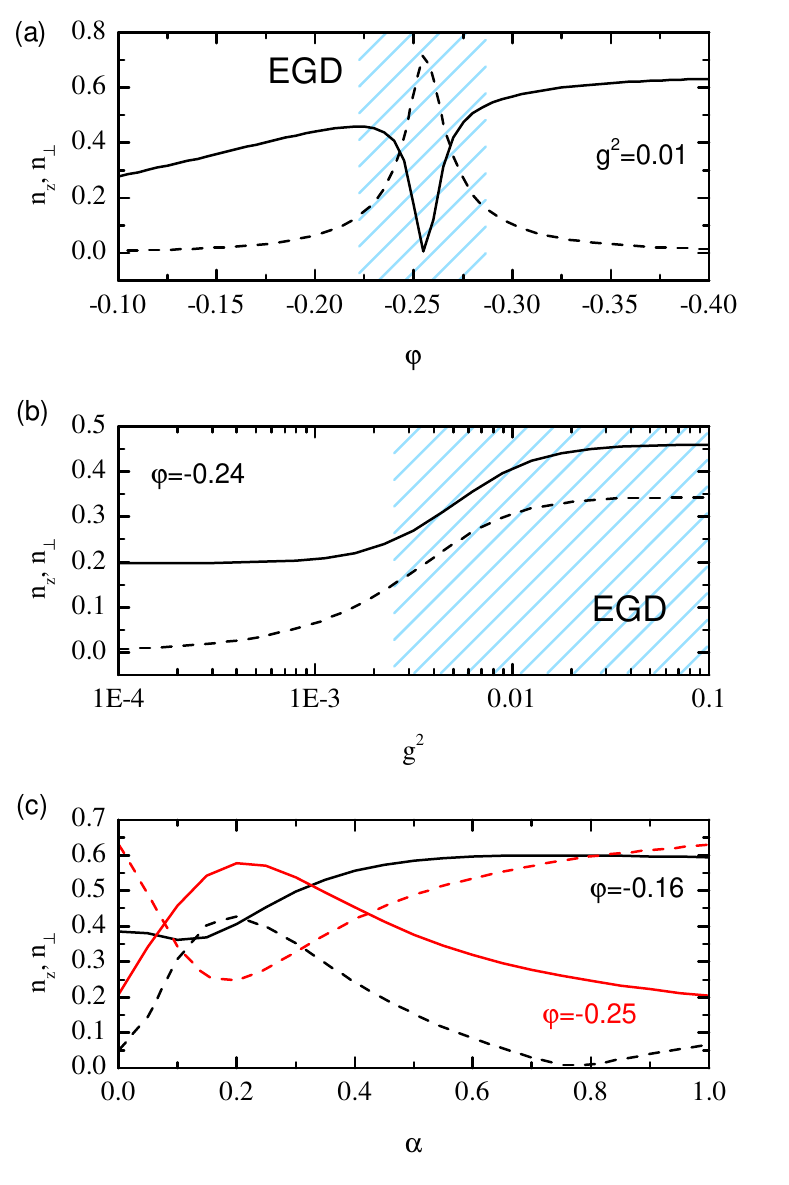}
   \end{center}
    \caption{\coloron SGD to EGD transition.
    $n_z$ (full lines) and $n_\perp$ (dashed lines) versus $\varphi$ for $A=\sigma_z$ and $g^2=0.01$ (a),
    versus $g^2$ for $A=\sigma_z$ and $\varphi=-0.24$ (b), and versus $\alpha$ for $\varphi=-0.16$ (black) and $\varphi=-0.25$ (red) and $g^2=0.1$ (c).
    EGD is attained is the highlighted regions of panels (a,b) and everywhere in panel (c).
    The drive parameters are the same as in \figref{fig:sluice}.}
    \label{fig:sgd2egd}
\end{figure}

In \figref{fig:sgd2egd}(a) we plot $n_z$ (full line) and $n_\perp$ (dashed line)
versus $\varphi$ in a neighborhood of $\varphi_c$. We choose the noise operator
$A=\sigma_z$ and $g^2=0.01$.
As $\varphi$ approaches $\varphi_c$ from either side, we witness the transition
from SGD to EGD, signalled by a revival of the coherence between Floquet states.
This revival is the first distinctive feature of EGD.
We then fix $\varphi$ to a value close to $\varphi_c$ and change $\chi$ by
changing $g^2$, to which $\chi$ is proportional.
The results are presented in \figref{fig:sgd2egd}(b).
For small values of $g^2$, the coherence between Floquet states is completely
lost ($n_\perp = 0$, SGD).
For larger values of $g^2$ (but still in the weak-coupling regime), $n_\perp$
takes a finite value (EGD).
Notice that as soon as either of the two limits is attained, the steady-state
populations do not depend on the exact value of $g^2$.

A second distinctive feature of EGD is that the relaxation basis
becomes strongly dependent on the type of noise.
To demonstrate this, we introduce a family of coupling operators
$A(\alpha)=\alpha \sigma_x + (1-\alpha)\sigma_z$, where $\alpha \in [0,1]$
parametrizes the angle between the reference basis of $H$ and the noise
operator.
In \figref{fig:sgd2egd}(c) we then plot $n_z$ (solid lines) and $n_\perp$
(dashed lines) versus $\alpha$ for two different values of $\varphi$. We set
$g^2=0.1$, a value ensuring that the EGD limit is attained for both cases.
Upon changing $\alpha$, $\rhost$ undergoes significant changes. This is in stark
contrast to what happens in SGD, where relaxation always takes place in the same
(Floquet) basis regardless of the coupling operator.

As $\rhost$ is determined by a set of algebraic equations,
obtained by setting $\dot \rho_{\alpha\beta} = 0$ in \eqref{eq:blum},
these features can be analytically addressed.
In the SGD limit, we obtain the same results as predicted by a
full secular approximation:
$\rho_{11}^{\rm st} = \rho_{11,{\rm SGD}}$, where $\rho_{11,{\rm SGD}}$
is a constant whose explicit expression is found in literature
\cite{Grifoni1998, Russomanno2011}, and $\rho_{12}^{\rm st} = O(\chi)$.
By contrast, in the EGD limit, $\rho_{11}^{\rm st} =
\rho_{11,{\rm EGD}}$ and $\rho_{12}^{\rm st} = \rho_{12,{\rm EGD}} + O(1/\chi)$.
The constants $\rho_{11,{\rm SGD}}$, $\rho_{11,{\rm
EGD}}$ and $\rho_{12,{\rm EGD}}$ are defined in \supplmat.

Altogether, the results of \figref{fig:sgd2egd} indicate that in
the EGD regime the environment is drastically influencing the dynamics of the
system. A measurement of $n_{x,y,z}$ can thus disclose
valuable information on both the type of noise and the strength of
system-environment coupling.

\mysec{Observation in a superconducting charge pump.}
We now discuss how to observe the transition in a superconducting
charge pump, the Cooper-pair sluice
\cite{Niskanen2003,Niskanen2005,Mottonen2008}.
\begin{figure}
    \begin{center}
   \includegraphics{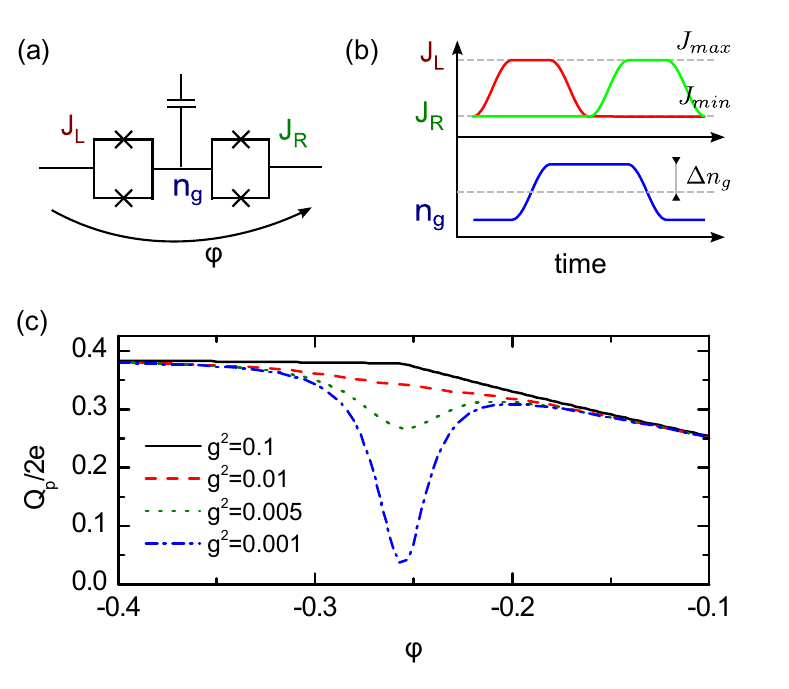}
   \end{center}
    \caption{Application to the Cooper-pair sluice.
    (a) Equivalent circuit of the sluice.
    (b) Time dependence of $J_L$, $J_R$ and $n_g$ during a pumping cycle.
    (c) Pumped charge $Q_p$ versus $\varphi$ for selected values of $g^2$.
    Relevant drive parameters are
    $\tau$=\SI{1}{ns}, $E_C$=\SI{1}{K}, $\Delta n_g=0.2$, $\Jmax=0.1E_C$, $\Jmin=10^{-3}\Jmax$.
    }
    \label{fig:sluice}
\end{figure}
The sluice, shown schematically in \figref{fig:sluice}(a), consists of a single
superconducting island, coupled to superconducting leads via two superconducting quantum
interference devices (SQUIDs). The SQUIDs are operated as Josephson junctions of
tunable energies $J_{L,R}(t)$. A gate electrode capacitively coupled to the
island provides a third control parameter by inducing a polarization charge
$n_g(t)$ in units of $2e$. The device is operated under a constant
superconducting phase bias $\varphi_S$.
In the charging regime $E_C\gg J_{L,R}$ ($E_C$ is the charging energy of the island),
the dynamics can be reduced to the
two lowest-energy charge states $|0\rangle $ and
$|1\rangle$ corresponding to zero and one excess Cooper pairs on the island,
respectively.
The system is then described by a pseudospin
Hamiltonian of the form discussed above, with effective field components $B_x(t)= \frac{1}{2}
J_+ (t)\cos\frac{\varphi}{2},\ B_y(t)= \frac{1}{2} J_-(t) \sin\frac{\varphi}{2}$,
and $B_z(t)= E_C \left[1/2 -n_g(t) \right]$, where $J_{\pm} (t) =
J_{L}(t) \pm J_{R}(t)$.
Pumping is achieved by steering the three parameters $J_L$, $J_R$ and $n_g$ in a periodic fashion,
as shown in \figref{fig:sluice}(b).

Starting from the given definitions, it can be shown that the 
pumping cycle of the sluice is a realization of the loop of \figref{fig:Floquet}(a),
with $\varphi=\varphi_S$.
The quasienergy gap of the sluice can thus be tuned by
changing the superconducting phase bias while performing exactly the same pulse
sequence.
This makes the sluice an excellent candidate to verify our theoretical
predictions through a direct measurement of the pumped charge $Q_p$
\cite{Mottonen2008,Gasparinetti2012}.

The main source of decoherence in the sluice is charge noise, due to
fluctuations in the gate voltage \cite{Pekola2010, Solinas2010}. We describe it
by putting $A=\sigma_z$, $\alpha=1$ and $g=C_g/C_\Sigma$, where $C_g$ and
$C_\Sigma$ are the gate-to-island capacitance and total island capacitance,
respectively.

In \figref{fig:sluice}(c) we plot $Q_p$ as a function of
$\varphi$ for different coupling strengths $g^2$. For small values of $g^2$,
corresponding to $\chi \ll 1$, $Q_p$ exhibits a dip around $\varphi_c$.
The dip appears in coincidence with the weakly avoided crossing in the
quasienergy spectrum [\figref{fig:Floquet}(b)], and stems from the mixing of
``adiabatic'' Floquet states at the crossing \cite{Hone2009,Russomanno2011}.
As $g^2$ is increased, an expanding neighborhood of $\varphi_c$ undergoes the
transition to EGD, producing an increase in $Q_p$.
Finally, for large enough values of $g^2$ the whole region of mixing is in the
EGD regime, and the dip has disappeared.
The observation of a finite pumped charge at the avoided quasienergy crossing
requires quantum coherence between Floquet states \footnote{As shown in
\cite{Russomanno2011}, the charge carried by a Floquet state is proportional to
the derivative of the corresponding quasienergy with respect to the phase bias
$\varphi$. At a symmetrically avoided quasienergy crossing, this implies that
the transferred charge vanishes for any incoherent superposition of the two
Floquet states.}; for this reason, it should be regarded as a direct signature
of EGD.

\mysec{Conclusions.}
A driven quantum system interacting with the environment exhibits a richer
scenario than an undriven one.
This is due to the emergence of an energy scale, the Floquet gap, that can
compete with decoherence rates in the vicinity of a quasienergy crossing.
This energy can be tuned by choosing the drive parameters.
We have identified two dynamical regimes and given an effective coupling parameter
governing the transition between the two. This transition manifests itself in
the quasistationary density matrix, in particular, in the revival of coherences
between Floquet states.
Finally, we have discussed how to observe the transition in a
superconducting charge pump. A closely related system that can be considered in
a similar spirit is the driven Cooper-pair box \cite{Wilson2007a,Wilson2010}.

While the SGD regime has been intensively studied, the EGD regime is vastly
unexplored. Due to its simplicity and applicability to a variety of different
systems, the present approach may emerge as a useful tool to study system-environment
interactions in open quantum systems.

\mysec{Acknowledgements.}
We would like to thank J.~Ankerhold, C.~Ciuti, V.~Gramich, M.~M\"ott\"onen, J.~Salmilehto
and M.~Silveri for useful discussions.
This work was supported by the Finnish National Graduate School in Nanoscience,
by the European Community's Seventh Framework Programme under Grant Agreement
No.~238345 ``GEOMDISS'', and partially by the Academy of Finland through its
Centres of Excellence Program (project No.~251748).
P.S. acknowledges financial support from 
FIRB -- Futuro in Ricerca 2012 under Grant No.~RBFR1236VV ``HybridNanoDev''.

\clearpage
\onecolumngrid
\appendix
 
\section{Quasistationary solution for the master equation}
\label{app:quasistationary_sol}

Starting from Eq.~\eqref{eq:blum} in the main text, we restrict ourselves to a
two-level system and set $\dot \rho_{\al\be} =0$.
We are not here interested in the details of the solution but only in its scaling with
respect to the quantities $g$, $\Omega$ and $\Fgap$.

We represent the
quasistationary solution $\rhost$ as a vector of components $(\rho_{11}^{\rm
st},\rho_{12}^{\rm st},\rho_{21}^{\rm st},\rho_{22}^{\rm st})$.
Using \eqref{eq:blum}, the definitions \eqref{eq:Rdef} and \eqref{eq:Gammas}
and the symmetries of the rates \eqref{eq:Gammas}, we find that
$\vec{\rho}_{\rm st}$ satisfies the following matrix equation:
\begin{equation} \label{eq:rhost_mat}
\left[
g^2 \Omega \left(
\begin{array}{cccc}
m & n & n^* & o \\
p & q & r & s \\
p^* & r^* & q & s^* \\
-m & -n & -n^* & -o
\end{array}
\right)
- i \Fgap \left(
\begin{array}{cccc}
0 & 0 & 0 & 0 \\
0 & 1 & 0 & 0 \\
0 & 0 & -1 & 0  \\
0 & 0 & 0 & 0
\end{array}
\right)
\right]
\cdot \vec{\rho}_{\rm st}
=0\ .
\end{equation}
where $m,\ldots, s$ are a set of dimensionless coefficients. We have factored
out a factor $g^2\Omega$ in the Redfield tensor \eqref{eq:Rdef}
as this is its expected scaling in the case we consider
(Ohmic environment, $\Omega \ll \omega_c$).

For simplicity, let us assume that all coefficients in \eqref{eq:rhost_mat} are real (it turns out that
this is often the case). The exact solution is then given by:
\begin{equation} \label{eq:rho_sol}
 \rho_{11}^{\rm st} = \frac{A}{C} \ , \rho_{12}^{\rm st} = \frac{D}{C} \ ,
\end{equation}
with
\begin{align*}
 A &=(q - r) [o (q + r) - 2 n s] g^4 \Omega^2 + o \Fgap^2 \ , \\
 C &= 2 n (q - r) (p - s) g^4 \Omega^2 - (m - o) [(q^2 - r^2)g^4 \Omega^2 + \Fgap^2)]\ , \\
 D &=  (m s - o p) [(q - r) g^2 \Omega + i \Fgap] g^2 \Omega \ .
\end{align*}

In the SGD regime $\chi \ll 1$, $A \approx o\Fgap^2$,
$C \approx (o-m)\Fgap^2$ and
$D \approx i (m s-o p) g^2 \Omega \Fgap$. An approximate solution is
\begin{subequations}\label{eq:rho_weak}
\begin{align}
 \rho_{11}^{\rm st} &= \frac{o}{o-m} \equiv \rho_{11,{\rm SGD}} \ , \label{eq:rho11_weak} \\
 \rho_{12}^{\rm st} &= i \frac{o p - m s}{m-o} \frac{g^2 \Omega}{\Fgap} ={\rm O}\left(\chi\right)
 \ , \label{eq:rho12_weak}
\end{align}
\end{subequations}
The coherence $\rho_{12}^{\rm st}$ vanishes for infinitesimally weak coupling, indicating
that the Floquet basis is the relaxation basis. Furthermore, 
\eqref{eq:rho11_weak} does not depend on the coupling strength.
These result are the same as obtained with a full secular approximation \cite{Grifoni1998,Russomanno2011}.

In the EGD regime $\chi \gtrsim 1$, an approximate solution is:
\begin{subequations}
\begin{align}
 \rho_{11}^{\rm st} &= \frac{o(q + r) - 2 n s}{2n(p - s)-(m - o)(q + r)} \equiv \rho_{11,{\rm EGD}} \ ,
 \label{eq:rho11_str} \\
 \rho_{12}^{\rm st} &= \frac{m s - o p }{2 n  (p - s) - (m - o) (q + r)}\left(
 1+\frac{i}{q-r} \frac{\Fgap}{g^2\Omega} \right) \equiv \rho_{12,{\rm EGD}}+{\rm O}\left(\frac1\chi\right) \ .
 \label{eq:rho12_str}
\end{align}
\end{subequations}
The coherence $\rho_{12}^{\rm st}$ in general does not vanish; in the
limit $\chi \gg 1$, it also approaches a constant value.

\end{document}